\documentclass[journal]{IEEEtran}
\ifCLASSINFOpdf
 \usepackage[pdftex]{graphicx}
\else
\fi
\usepackage[cmex10]{amsmath}
\usepackage{color}

\begin{document}
%
% paper title
% can use linebreaks \\ within to get better formatting as desired
% Do not put math or special symbols in the title.
\title{Electronic Implementation of the Mackey-Glass Delayed Model}
%
%
% author names and IEEE memberships
% note positions of commas and nonbreaking spaces ( ~ ) LaTeX will not break
% a structure at a ~ so this keeps an author's name from being broken across
% two lines.
% use \thanks{} to gain access to the first footnote area
% a separate \thanks must be used for each paragraph as LaTeX2e's \thanks
% was not built to handle multiple paragraphs
%

\author{Pablo Amil, Cecilia Cabeza, Arturo   C. Mart\'{i}% <-this % stops a space 
 
\thanks{Pablo Amil, Cecilia
  Cabeza, and Arturo C. Mart\'{\i} are with the Physics Institute,
  Universidad de la Rep\'ublica, Igu\'a 4225, Montevideo, Uruguay,
  (see http://www.fisica.edu.uy).}% <-this % stops a space 
  \thanks{Manuscript received ------,
2014; revised ------, 2014.}}

\markboth{Transactions on Circuits and Systems I,~Vol.~X, No.~Y, -----}%
{Shell \MakeLowercase{\textit{et al.}}: Bare Demo of IEEEtran.cls for Journals}
% The only time the second header will appear is for the odd numbered pages
% after the title page when using the twoside option.
% 
% *** Note that you probably will NOT want to include the author's ***
% *** name in the headers of peer review papers.                   ***
% You can use \ifCLASSOPTIONpeerreview for conditional compilation here if
% you desire.

% If you want to put a publisher's ID mark on the page you can do it like
% this:
%\IEEEpubid{0000--0000/00\$00.00~\copyright~2012 IEEE}
% Remember, if you use this you must call \IEEEpubidadjcol in the second
% column for its text to clear the IEEEpubid mark.

% use for special paper notices
%\IEEEspecialpapernotice{(Invited Paper)}

% make the title area
\maketitle

% As a general rule, do not put math, special symbols or citations
% in the abstract or keywords.
\begin{abstract}
The celebrated Mackey-Glass model describes the dynamics of
physiological \textit{delayed} systems in which the actual evolution
depends on the values of the variables at some \textit{previous}
times. This kind of systems are usually expressed by delayed
differential equations which turn out to be infinite-dimensional. In this contribution, 
an electronic implementation mimicking the Mackey-Glass model is proposed. 
New approaches for both the nonlinear function and the delay block are made. Explicit equations for the actual
evolution of the implementation are derived. Simulations of the
original equation, the circuit equation, and experimental data show great concordance.
\end{abstract}

% Note that keywords are not normally used for peerreview papers.

% For peer review papers, you can put extra information on the cover
% page as needed:
% \ifCLASSOPTIONpeerreview
% \begin{center} \bfseries EDICS Category: 3-BBND \end{center}
% \fi
%
% For peerreview papers, this IEEEtran command inserts a page break and
% creates the second title. It will be ignored for other modes.
\IEEEpeerreviewmaketitle

\section{Introduction}

In 1977, a paper entitled \emph{Oscillation and Chaos in Physiological
  Control Systems} \cite{mackey1977oscillation} was published in
\textit{Science}.  This paper, by Michael C. Mackey and Leon Glass (MG),  dealt with physiological processes,
mainly respiratory and hematopoietic (\textit{i.e.} formation of blood cellular
components) diseases in which time delays play a significant role. In
effect, in the production of blood cells there is a considerable delay
between the initiation of cellular production in the bone narrow and
the release into the blood. In general, in these processes, the evolution of the
system at a given time not only depends on the state of the system at
the current time but also on the state of the system at
\textit{previous} times.  The MG's work had an impressive
impact.  Since its publication, it exhibits nearly 3000 cites in
scientific journals and, at present, Google reports more than two
millions results for the search \textit{Mackey-Glass}.

In their pioneering paper, Mackey and Glass  showed that a variety
of physiological systems can be adequately described in terms of
simple nonlinear delay-differential equations. The model proposed by MG
exhibits a wide range of behaviors including periodic or chaotic
solutions.  The importance of the MG model lies in the fact that the onset of
some diseases are associated with alterations in the periodicity of
certain physiological variables, for example, irregular breathing
patterns or fluctuations in peripheral blood cell counts.

The dynamics of processes involving time delays as those studied by MG  is far more complex than non-delayed, \textit{i.e.}, instantaneous
systems. Actually, if the dynamics of a  system  at time $t$ depends on the state of the system at time $t-\tau$, 
the information needed to predict the evolution is content in the whole interval $(t-\tau,t)$.  Thus, the evolution of
a delayed system depends on \textit{infinite} previous values of the variables.
From the mathematical point of view, delayed systems are modelled in
terms of delayed differential equations (DDEs) and one single DDE is
equivalent to infinite ordinary differential equations (ODEs). Due to their infinite dimensionality, the
accuracy of numerical simulations of DDEs is specially delicate.  In
practice, this problem is avoided considering large transients. However, there persist doubts about the stability and
accuracy of the methods used to numerically integrate DDEs.

Thank to its richness in behaviors, the Mackey-Glass model has acquired relevance
of its own \cite{junges2012intricate,shahverdiev2006chaos,berezansky2006mackey,PhysRevE.75.016207,namajunas1995stabilization}.
One direct application is using MG model as a simple way to generate a chaotic signal \cite{grassberger1983measuring,grassberger1983estimation}
to be used in multiple ways as for example to check methods to characterize chaotic  
measures or stability schemes \cite{namajunas1995electronic,namajunas1995stabilization}.

Several electronic implementations of MG model have been proposed
\cite{kittel1998generalized,namajunas1995electronic,namajunas1995stabilization,wan2009bifurcation,tateno2012nonlinear}.
In Ref.~\cite{namajunas1995electronic}, an electronic implementation
based on an \textit{analog delay line} was proposed to address the
problem of controlling high dimensional chaos in infinite dimensional
system.  The focus was on the stabilization of unstable steady states
(USS) in a electronic analog to the MG system.  A possible application
of the MG model is to employ several delayed values of the variable
$x(t-\tau_1)$, $x(t-\tau_2)$, $x(t-\tau_3)$....  instead of only
one \cite{PhysRevE.75.016207,tateno2012nonlinear}.
More specifically, Taneto and Uchida \cite{tateno2012nonlinear}, investigated the
generation of chaos in a Mackey-Glass electronic circuit with two
time-delayed feedback loops observing different dynamical behaviors
when the two delay times were changed. The ratio of the two time
delays was found crucial to enhance or suppress the chaotic
dynamics. The synchronization of chaos in unidirectionally-coupled
Mackey- Glass electronic circuits with two time delays was also
investigated confirming that synchronization of chaos can be achieved
even in the presence of the two time- delayed feedback
loops. High-quality synchronization of chaos can be achieved at the
strong coupling strengths and parameter matching conditions between
the two circuits.

% Let us consider the production of blood cells, consider a homogenous population of  mature circulating cells of density $P$. 
% There is a significant delay $\tau$ between the initiation of celullar production  in the bone narrow and the release into the blood. 
% As seen in Ref.~\cite{mackey1977oscillation}  (Eq.~4b), the delay differential equation is
% \begin{equation}
% \frac{dP}{dt}=\frac{\beta_{0}\Theta^{n}P_{\tau}}{\Theta^{n}+P_{\tau}^{n}}-\gamma P
% \label{eq1}
% \end{equation}
% Where $P_{\tau}\left(t\right)=P\left(t-\tau\right)$. $\beta_{0}$, $\gamma$,
% $\Theta$, and $n$ are model parameters.

The goal of this paper is to propose a novel electronic implementation
of the Mackey-Glass system.  We show how this implementation
approximates the original equation by deriving its evolution equation.
Then we show results of simulations comparing the original
Mackey-Glass equation and the effective circuit equation with the
experimental data, which all show great agreement between each other.

This paper is organized as follows. Section~\ref{sec:EI} deals
specifically with the design of the circuit. We pay special attention to the delay block and the function block which 
implement the nonlinear term. In Sec. \ref{sec:EE} we derive the governing equation of the designed
circuit.  For this circuit, in Sec.~\ref{sec:S&A} we present both, numerical and
experimental results.  Finally, in Sec.~\ref{sec:con} we draw our conclusions.

\section{The Model and its Electronic Implementation} 
\label{sec:EI}

Let us consider the production of blood cells. The homogeneous density
of a population of mature circulating cells is denoted $P$.  As mentioned before, there	 is
a significant delay $\tau$ between the initiation of cellular
production in the bone narrow and the release into the blood.  According to Ref.~\cite{mackey1977oscillation} (Eq.~4b), 
the delay differential equation governing the evolution of the population is
\begin{equation}
\frac{d	P}{dt}=\frac{\beta_{0}\Theta^{n}P_{\tau}}{\Theta^{n}+P_{\tau}^{n}}-\gamma P,
\label{eq1}
\end{equation}
where $P_{\tau}\left(t\right)=P\left(t-\tau\right)$. The parameters $\beta_{0}$, 
$\Theta$, and $n$ are related to the production rate while  $\gamma$ determines the decay rate of the cells.

To simplify the Eq.~\ref{eq1}, we will reduce it to an equivalent
equation with less model parameters. To that end, we will define a new
state-variable and a new independent-variable as $x = P/{\Theta}$ and
$ t' = \gamma t$ respectively.  We also define  new  parameters such that
$\Gamma = \gamma \tau$ and $\alpha = \beta_{0}/ \gamma$.  Thus, the
Eq.~\ref{eq1} can be now written as
\begin{equation}
\frac{dx}{dt'}=\alpha\frac{x_{\Gamma}}{1+x_{\Gamma}^{n}}-x
\label{eq:MGlinda}
\end{equation}
being $x_{\Gamma}(t')=x(t'-\Gamma)$. The electronic implementation
will mimic this equation by properly setting $\alpha$ and $\Gamma$.

Concerning the dynamics of the MG model given by Eqs.~\ref{eq1} or
\ref{eq:MGlinda}, the role of the time delay $\tau$ is crucial.  For
the instantaneous system, $\tau=0$, depending on the parameter values,
there is only one positive stable fixed point.  However, as $\tau$, is
increased, the initially stable equilibrium point becomes unstable and
periodic solutions appear \cite{namajunas1995stabilization}.  If
$\tau$ is further increased a sequence of bifurcations gives place to
oscillations with higher periods and aperiodic behavior. The value of
$n$ considered in \cite{mackey1977oscillation} was $n=10$, but similar
behaviors can be seen with other values of $n$
\cite{junges2012intricate}.

\subsection{Block diagram}
\begin{figure}
\begin{centering}
\includegraphics[width=0.95\columnwidth]{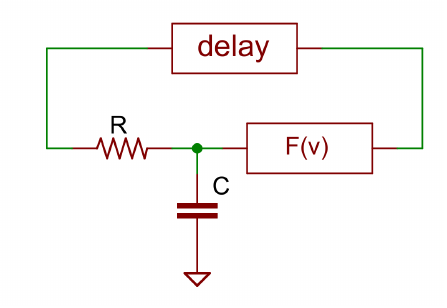}
\par\end{centering}
\caption{Block schematic of the complete circuit: delay and nonlinear function blocks, with the RC}
\label{fig:MGbloc}
\end{figure}

The electronic implementation was divided in two main parts: the delay
block, which presents only a time shift between its input and its
output; and the function block, which implements the nonlinear term of
the equation. After having these blocks the complete circuit looks as
in Fig. \ref{fig:MGbloc}. In this scheme the function block implements
the first term in the r.h.s. of Eq.~\ref{eq1} or Eq.~\ref{eq:MGlinda}
without delay
\begin{equation}
f\left(v\right)=\beta\frac{v}{\theta^{n}+v^{n}}\label{eq:fx},
\end{equation}
and the delay block implements the transfer function:
\begin{equation}
v_{out}\left(t\right)=v_{in}\left(t-\tau\right).
\label{eq:transfer}
\end{equation}
Assuming ideal behavior of both blocks, the equation for the potential at the capacitor terminals is given by 
\begin{equation}
\frac{dv_{c} \left(t\right)}{dt}=\frac{1}{RC}\left[f\left(v_{c}\left(t-\tau\right)\right)-v_{c}\left(t\right)\right]\label{eq:ElecEq}
\end{equation}
which can be identify with Eq.~\ref{eq:MGlinda} by setting
$t'={t}/{RC}$,
$x={v_{c}}/ {\theta}$,
$\Gamma={\tau} / {RC}$, and 
$\alpha={\beta}/ {\theta^{n}}$.

\subsection{Delay block}
\begin{figure}
\begin{centering}
\includegraphics[width=0.98\columnwidth]{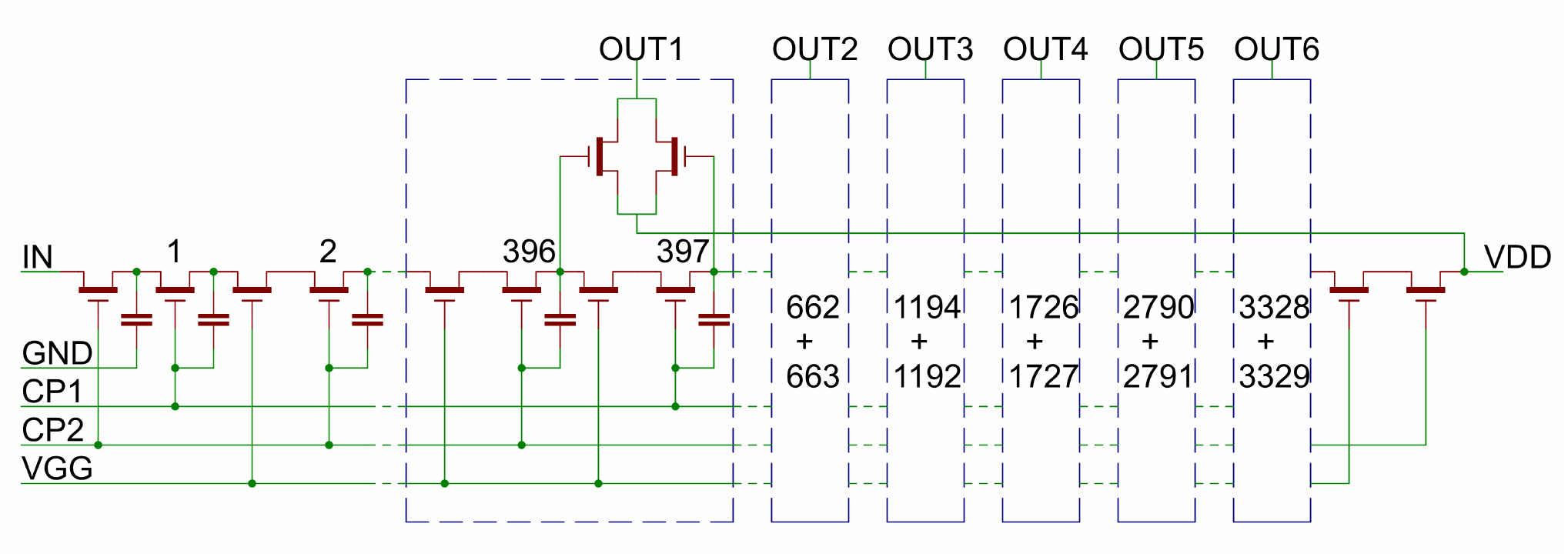}
\end{centering}
\caption{\label{fig:BBD}Internal structure of a Bucket Brigade Device (BBD)}
\end{figure}

The purpose of the delay block is to copy the input to the output after
some time delay.  The implementation of this block with analog
electronic is possible using a Bucket Brigade Device (BBD), which is a
discrete-time analog device, internally it consists of an array of $N$
capacitors in which the signal travels one step at a time, as shown in
Fig. \ref{fig:BBD}. In our implementation we used the integrated
circuits MN3011 and MN3101 as BBD and clock signal generator
respectively.

\begin{figure}
\begin{centering}
\includegraphics[width=0.98\columnwidth]{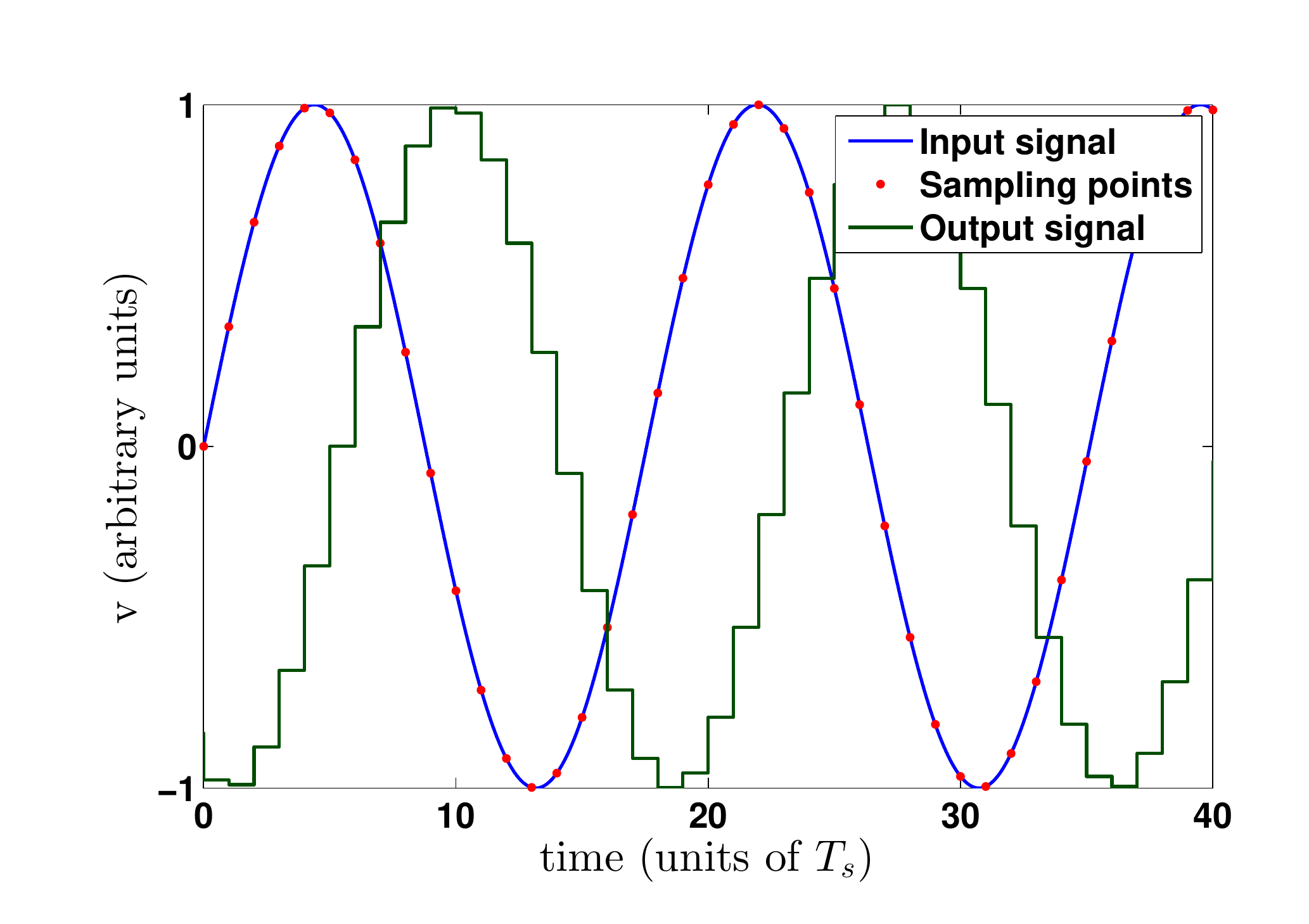}
\par\end{centering}
\caption{\label{fig:BBDEx}Simulated example  of  input an output signals of a BBD and sampling points. For the sake of clarity,
the offset and gain are null and $N=5$.}
\end{figure}

This approach of implementing a delay approximates the desired
transfer function given by Eq.~\ref{eq:transfer}, by sampling the
input signal and outputting those samples $N$ clock periods later. The
effective transfer equation is the following
\begin{equation}
v_{out}\left(t\right)=g_{d} v_{in}\left(T_{s} \left\lfloor  \frac{t}{T_{s}}-N+1 \right\rfloor  \right)+V_{d} 
\end{equation}
where $T_{s}$ is the sampling period of the BBD, and $N$ is the number
of capacitors in the array, $g_{d}$ and $V_{d}$ stand for gain and
offset voltage introduced by the BBD respectively. The symbols
$\lfloor .\rfloor$ stand for the integer part.  In the MN3011 the
sampling period can vary between $5\mu s$ and $50\mu s$.  The number
of capacitors in the array, $N$, can be selected among the values
provided by the manufacturer ($N=396, 662, 1194, 1726, 2790,
3328$). An example of its functionality can be seen in
Fig. \ref{fig:BBDEx} where, to fully appreciate the input and output
signals, the offset and the gain are null and a simulated value of $N$
(non provided by the manufacturer) is employed.

\begin{figure}
\begin{centering}
\includegraphics[width=0.98\columnwidth]{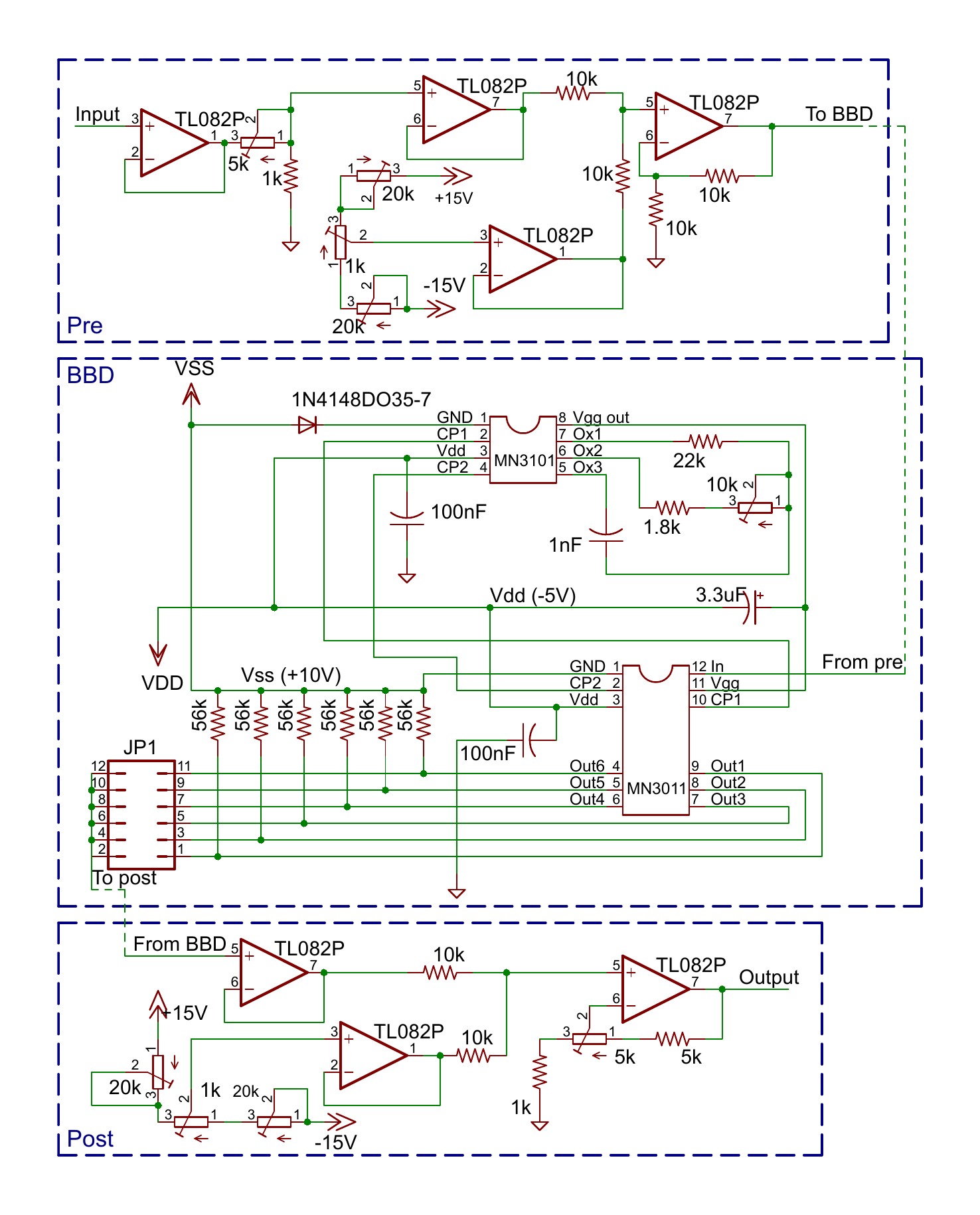}
\end{centering}
\caption{\label{fig:Retardo} Circuit schematic of the delay block. The blue-dashed boxes correspond to the pre-amplification, BBD and post-amplification.}
\end{figure}

\begin{figure}
\begin{centering}
\includegraphics[width=0.98\columnwidth]{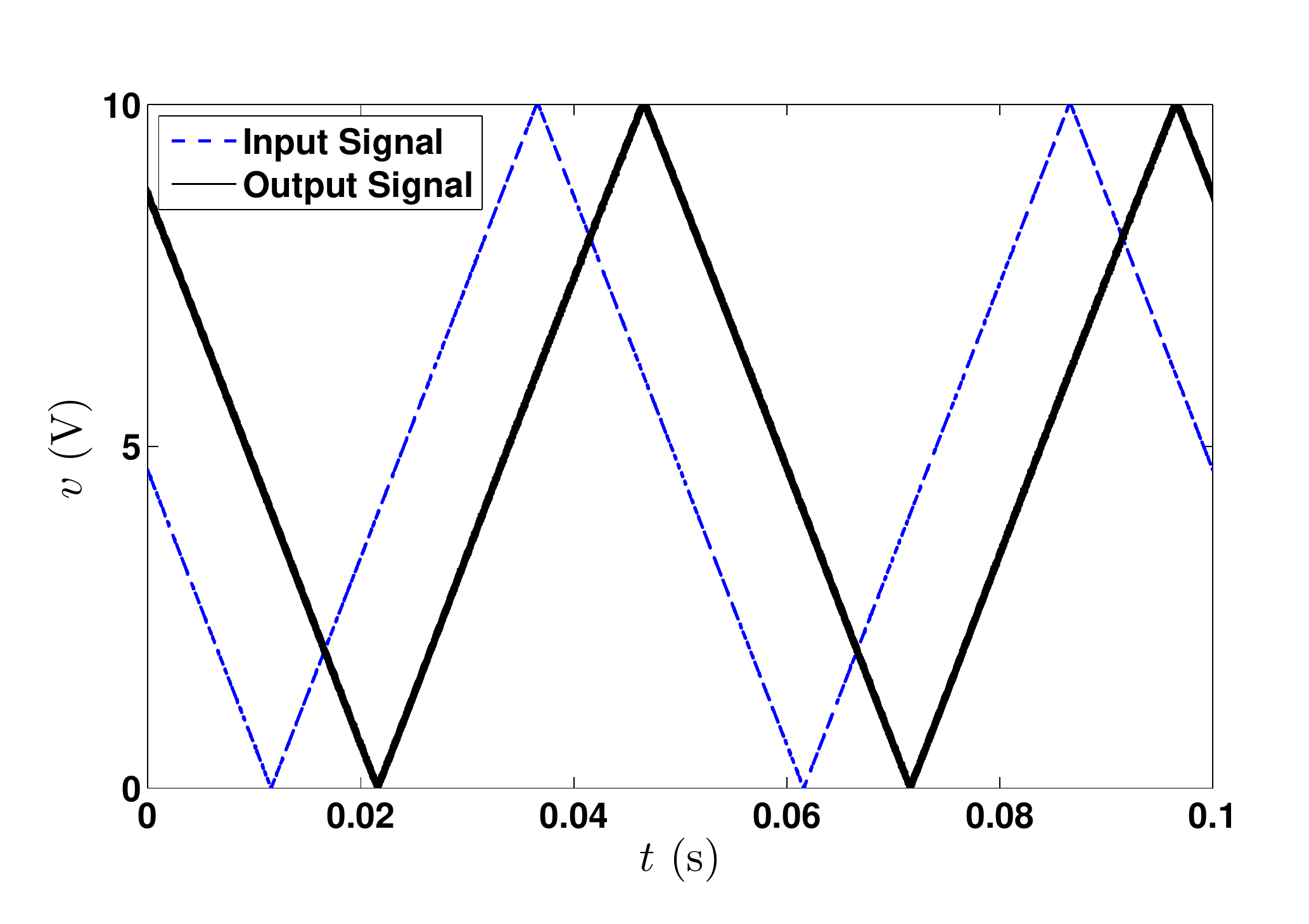}
\par\end{centering}
\caption{\label{fig:BBDExEx}Test signals of the delay block. Input signal
(dashed line), and output signal (full line).}
\end{figure}

In order to avoid the effects of offset and gain, and also to expand the intrinsic
dynamic range of the BBD (originally between $0$ V and $4$ V), the complete delay block included pre- and post-amplification and
offset adjust as shown in Fig.~\ref{fig:Retardo}. The delay was set to $10$~ms using  $N=1194$. All trimmers in the circuit were adjusted to obtain
a dynamic range  between $0$~V and $10$~V, and a transfer equation as follows
\begin{equation}
v_{out}\left(t\right)=v_{in}\left(T_{s}\left\lfloor   \frac{t}{T_{s}}-N+1 \right\rfloor  \right).
\label{eq:DelayBien}
\end{equation}
Input and output test signals are shown in Fig. \ref{fig:BBDExEx}.

It is worth noting  that, due to its internal clock,  a certain amount  of  high-frequency noise is introduced by the
delay block. This is the reason that explains the ordering of the blocks in which  the delay block is after the function block and 
inmediately before the  RC circuit.  With this configuration,  most of this noise is supressed.

\subsection{Function block}
\begin{figure}
\begin{centering}
\includegraphics[width=0.98\columnwidth]{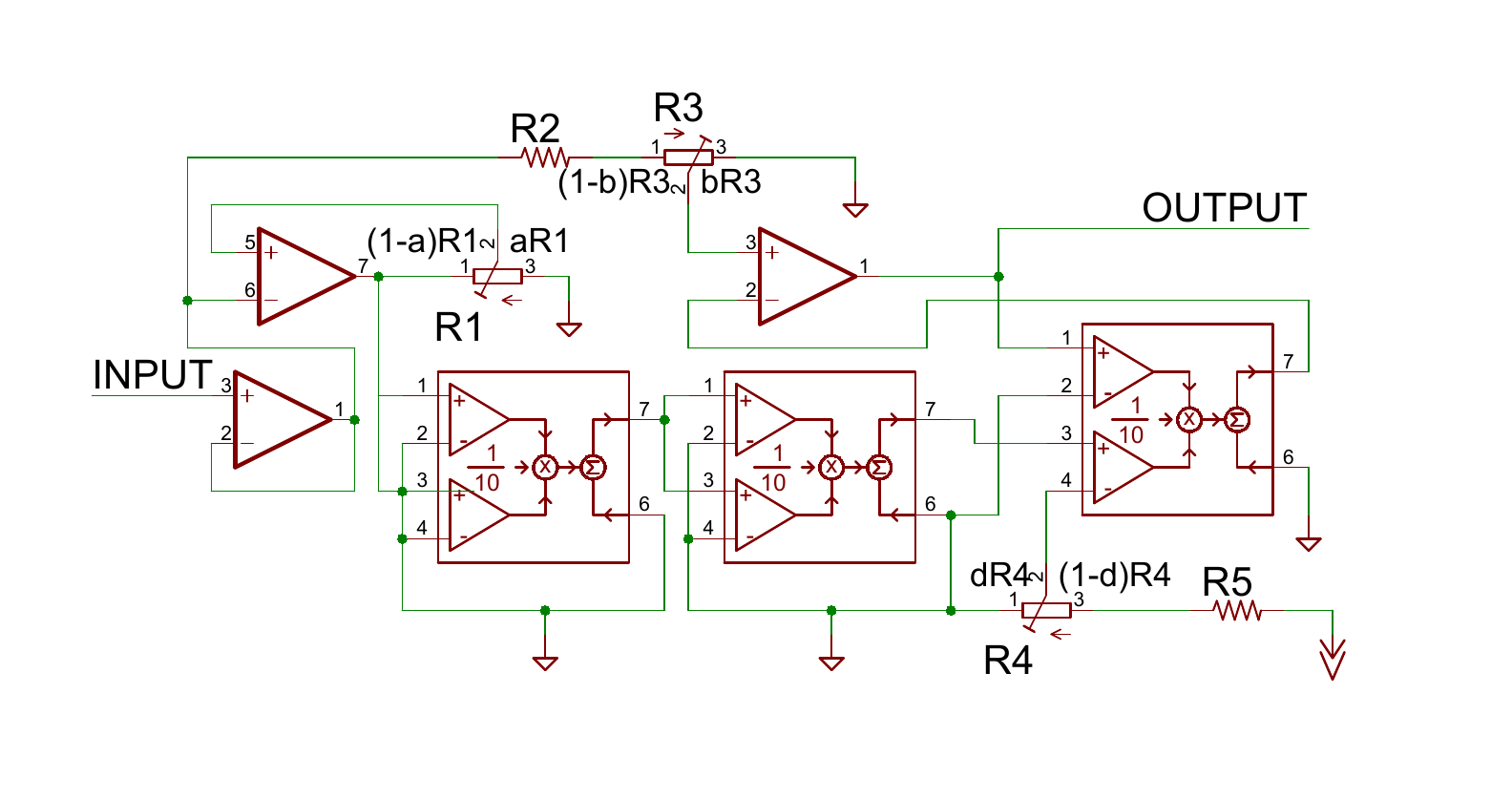}
\par\end{centering}
\caption{\label{fig:Fx}Circuit schematics of the nonlinear function. 
$R_{1}=20$ k$\Omega $, $R_{2}=56$k$\Omega $, $R_{3}=20$k$\Omega $, $R_{4}=2$k$\Omega $, $R_{5}=56$k$\Omega $}
\end{figure}

The function block was implemented for $n=4$, as shown in the circuit
schematics in Fig. \ref{fig:Fx}. Other values of the exponent $n$
could be similarly implemented.  Assuming all the operational
amplifiers are working in the linear region, the transfer function
results as in Eq.~\ref{eq:fx} by defining $\theta= k_1 a d^{\frac14} $
and $\beta= k_2 b^{\frac 14} $ where $k_1$ and $k_2$ are constants
that depend on the circuit and $a$, $b$ and $d$ depend on the
positions of the trimmers indicated in Fig.~\ref{fig:Fx}	.  The parameter $a$ does
not depend on the dimensionless parameters that determine the evolution
of the system, $\alpha$ and $\Gamma$, but only depends on $\theta$
that sets the amplitude of the oscillations. Therefore $a$ can be set
to avoid all kind of saturation in the circuit without changing the
dynamical properties of the oscillations.

% \begin{equation}
% \left\{ \begin{array}{c}
% v_{out}=b\frac{R_{3}}{R_{3}+R_{2}}a^{4}10^{4}\frac{v_{in}}{v_{in}^{4}+v_{cc}10^{3}a^{4}\frac{R_{4}}{R_{4}+R_{5}}c}
% \end{array}\right.\label{eq:10}
% \end{equation}
% which can become eq. \ref{eq:fx} by setting

\begin{figure}
\begin{centering}
\includegraphics[width=0.98\columnwidth]{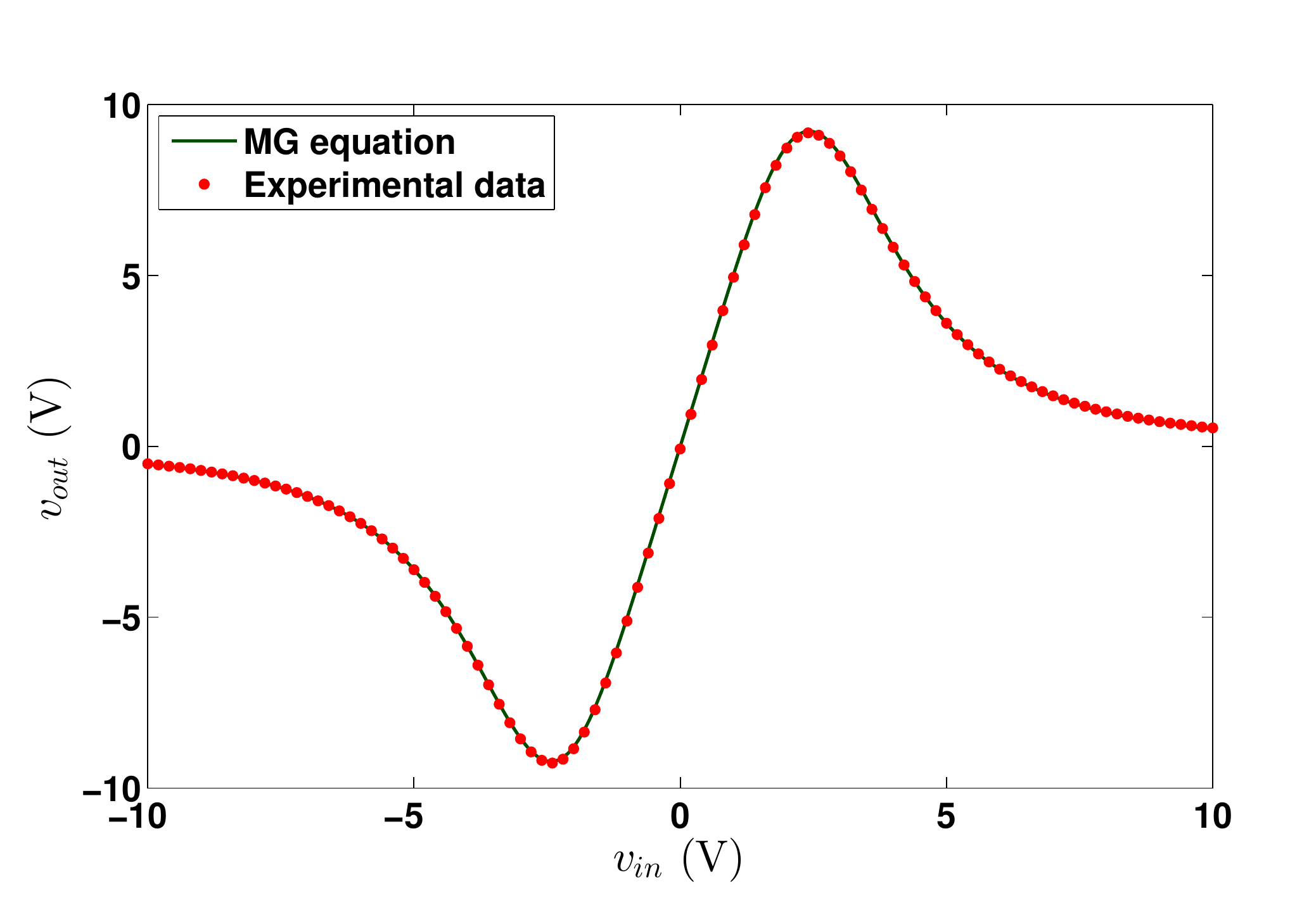}
\caption{\label{fig:FxGra}Input-output of the function block, experimental data (red dots) and
Eq.~\ref{eq:fx} (green line). Parameter values: $\beta=525V^{4}$ and  $\theta=3.19V$}. 
\end{centering}
\end{figure}

Integrated circuits AD633JN  and AD712JN  were used  to implement sums, multiplications and divisions
because of their simplicity, accuracy, low noise and low offset voltage. In Fig. \ref{fig:FxGra},
an input-output graph  of this block is depicted and  compared with Eq.~\ref{eq:fx}.
Although it was tested from $-10$ V to $10$ V, it is used only with positive voltages as they  are the only relevant in the MG model.

\section{Effective equation}
\label{sec:EE}
Since the delay block we used approximates an ideal delay, the
effective equation of the implemented circuit will also approximate
the original Mackey-Glass equation. By means of the Kirchof's law
applied to the entire circuit shown in Fig.\ref{fig:MGbloc} we obtain
\begin{equation}
\frac{dv_{c}(t) }{ dt  } =
\frac{1}{RC}\left[f\left(v_{c}\left(T_{s}\left\lfloor \frac{t}{T_{s}}-N+1\right\rfloor \right)\right)-v_{c}\right]\label{eq:circuit1}.
\end{equation}

The output of the delay block remains constant in each clock period,
so it seems natural to solve Eq. \ref{eq:circuit1} in steps.  Let us
solve it, then, for $jT_{s}\leq t<\left(j+1\right)T_{s}$ and let
$v_{i}=v_{c}\left(iT_{s}\right)$, then
\begin{equation}
\frac{dv_{c}}{dt}\left(t\right)=\frac{1}{RC}\left[f\left(v_{j-N+1}\right)-v_{c}\right].
\end{equation}
Since $f\left(v_{n-N+1}\right)$ is constant, this equation can be readily solved, 
the solution knowing the value of $v_{c}$ in $t=jT_{s}$ is
\begin{equation}
v_{c}\left(t\right)=\left(v_{j}-f\left(v_{j-N+1}\right)\right)e^{\frac{nT_{s}-t}{RC}}+f\left(v_{j-N+1}\right).
\end{equation}
Setting $t=(j+1) T_s$ and substituting the expression for $f(v)$ given in Eq.~\ref{eq:fx} it results
% \begin{equation}
% v_{j+1}= \left(v_{j}-f\left(v_{j-N+1}\right)\right)e^{\frac{-T_{s}}{RC}}+f\left(v_{j-N+1}\right).
%  \end{equation}
% 
% 
% Using the relations in eq. \ref{eq:setadim}, with $\tau=NT_{s}$, Eq.
% \ref{eq:iterfea} becomes
\begin{equation}
v_{j+1}=   v_{j}e^{\frac{-T_{s}}{RC}}+ (1 - e^{\frac{-T_{s}}{RC}})   \beta\frac{v_{j-N+1}}{\theta^{n}+v_{j-N+1}^{n}}.
%v_{j+1}=\left(v_{j}-\beta\frac{v_{j}}{\theta^{n}+v_{j}^{n}}\right)e^{\frac{-T_{s}}{RC}}+\beta\frac{v_{j}}{\theta^{n}+v_{j}^{n}}\label{eq:iterfea}
\label{eq:iterlinda}
\end{equation}
% \begin{equation}
% x_{j+1}=x_{j}e^{-\nicefrac{\Gamma}{N}}+\left(1-e^{-\nicefrac{\Gamma}{N}}\right)\alpha\frac{x_{j-N+1}}{1+x_{j-N+1}^{n}}\label{eq:iterlinda}
% \end{equation}
It can be easily seen that this discrete time effective equation
approaches to the original continuous time equation,
Eq.~\ref{eq:ElecEq}, when $N$ grows to infinity.

\section{Simulations and results}
\label{sec:S&A}
To compare the solutions of the original (continuous-time) model with
the effective (discrete-time) model, we performed simulations of both
with the same parameters values.  The original equation was simulated
using a 5th order Runge-Kutta scheme with variable time step.  Since
$N$ tending to infinity would reconstruct the original equation, the
effective equation was simulated using the less favorable case,
$N=396$, and assumed that other values of $N$ perform better than this
value, For the circuit we used $N=1194$, where we measured the voltage
immediately after the delay block, and reconstructed the voltage in
the capacitor with a tuned digital filter.

% 
% With this simulations, bifurcation diagrams were performed Figs. \ref{fig:C-TBif}
% and \ref{fig:D-TBif} showing great agreement.
% 
% ($n$,
% $\Gamma$, and $\alpha$) with both: the original equation using
% Runge-Kutta method of order 4-5 with variable time step and the effective
% equation; and performed measurements on the circuit. We set $N$ to
% the worst case, $N=396$ in the effective equation simulations and
% assume that other values of $N$ will behave as well as this, since
% $N$ tending to infinity would reconstruct the original equation,
% while we used $N=1194$ for the circuit, where we measured the the voltage 
% immediately after the delay block, and reconstructed the voltage in the  capacitor with a tuned digital filter.

% In all three cases, the parameter $\Gamma$ was varied to perform
% bifurcation diagrams of the local maximuns. Such diagrams are shown in Figs. \ref{fig:C-TBif}
% and \ref{fig:D-TBif} respectively for simulations of original equation
% and effective equation, while experimental results are shown in Fig.
% \ref{fig:E-TBif}, all three of them showing great agreement between each other.

\begin{figure}
\begin{centering}
\includegraphics[width=0.98\columnwidth]{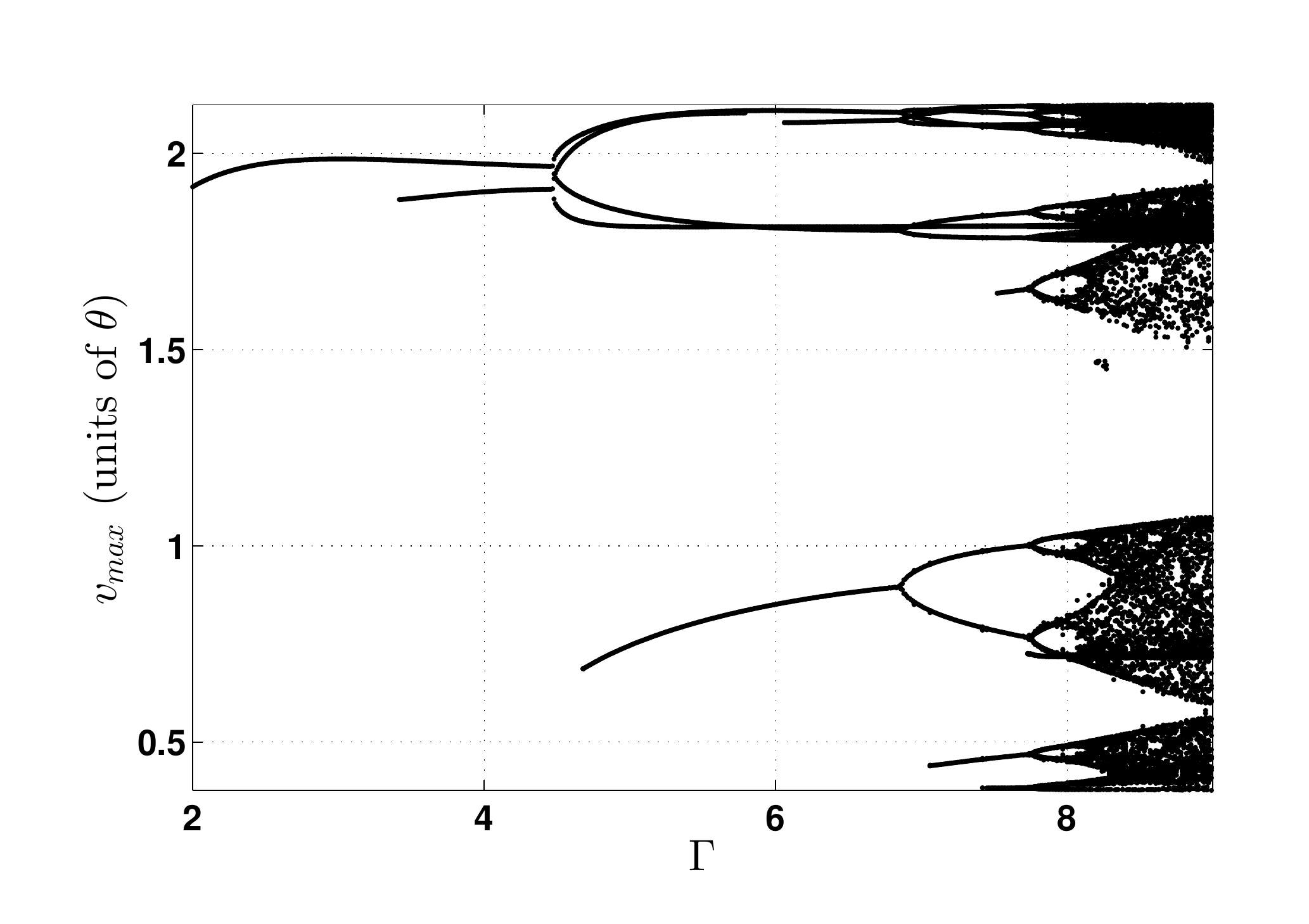}
\par\end{centering}
\caption{\label{fig:C-TBif}Bifurcation diagram as a function of the  time  delay corresponding to the continuous-time simulations.  
Parameter  values: $n=4$ and $\alpha=3.73$.}
\end{figure}

\begin{figure}
\begin{centering}
\includegraphics[width=0.98\columnwidth]{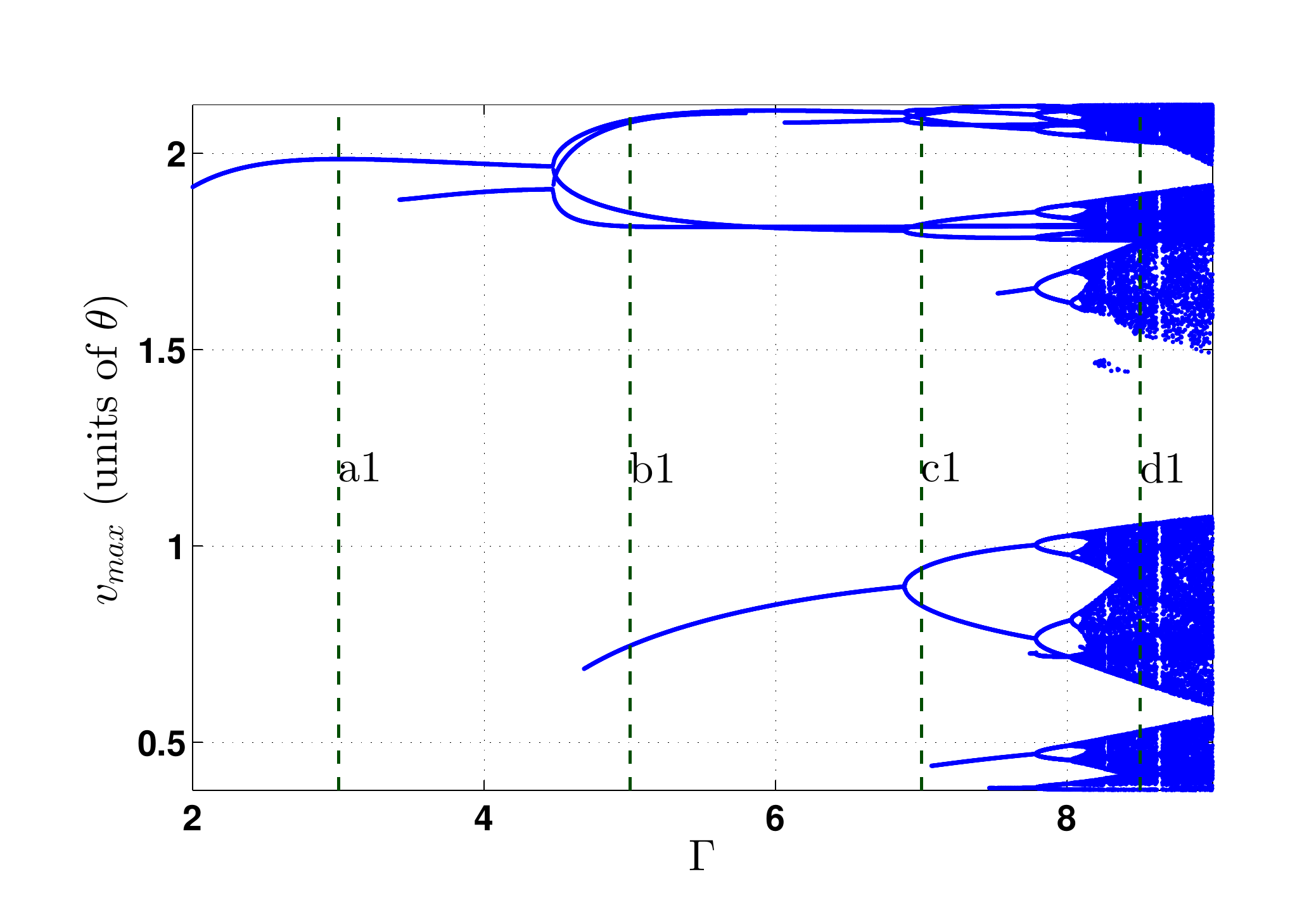}
\par\end{centering}
\caption{\label{fig:D-TBif}Discrete time version of the bifurcation  diagram shown in Fig.\ref{fig:C-TBif} with $N=396$. The vertical
  lines labeled with a1, b1, c1 and d1 correspond to the temporal   series shown in Figs.~\ref{fig:Series1} and \ref{fig:Series2}.}
\end{figure}

\begin{figure}
\begin{centering}
\includegraphics[width=0.98\columnwidth]{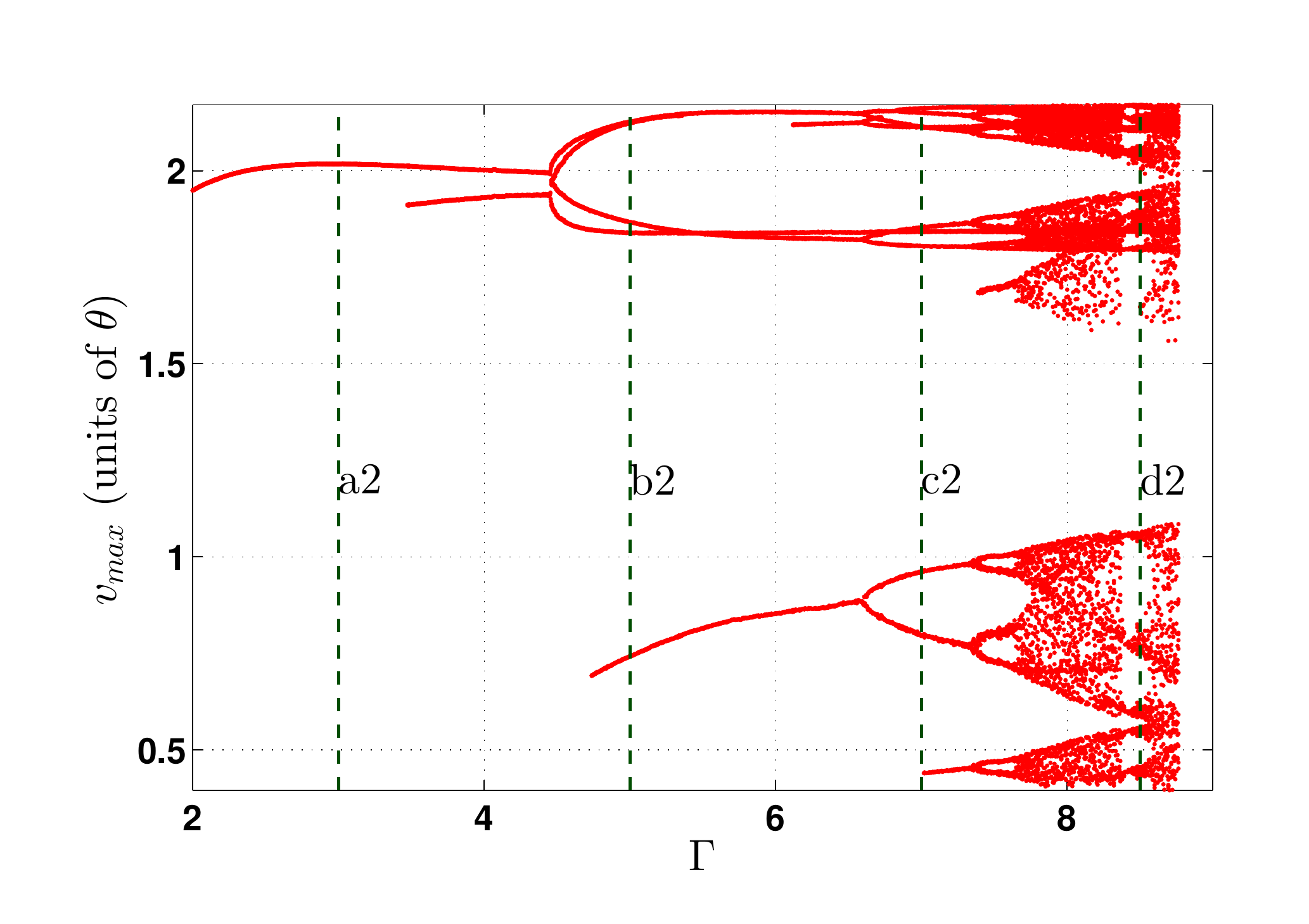}
\par\end{centering}
\caption{\label{fig:E-TBif}Experimental results $n=4$, $\alpha=3.73$   and $N=1194$. The labels a2, b2, c2 and d2 correspond to the
  temporal series shown in Figs.~\ref{fig:Series1} and   \ref{fig:Series2}.}
\end{figure}

Bifurcation diagrams were obtained by plotting the maximum of temporal
series, as a function of $\Gamma$, which is $\tau$ in units of $RC$.
Such diagrams are shown in Figs. \ref{fig:C-TBif} and \ref{fig:D-TBif}
respectively for simulations of original equation and effective
equation, while experimental results are shown in
Fig. \ref{fig:E-TBif}. We can observe great agreement in all cases.
The three figures present a typical behavior of DDEs: singles branches that appear (\textit{out of the blue}) or disappear
for certain control parameter values. In addition, the familiar, already present in ODEs, period-doubling branches and chaotic
behavior also emerge as the control parameter $\Gamma$ is increased. These results are in agreement with previous works
\cite{junges2012intricate}.

\begin{figure}
\includegraphics[width=0.98\columnwidth]{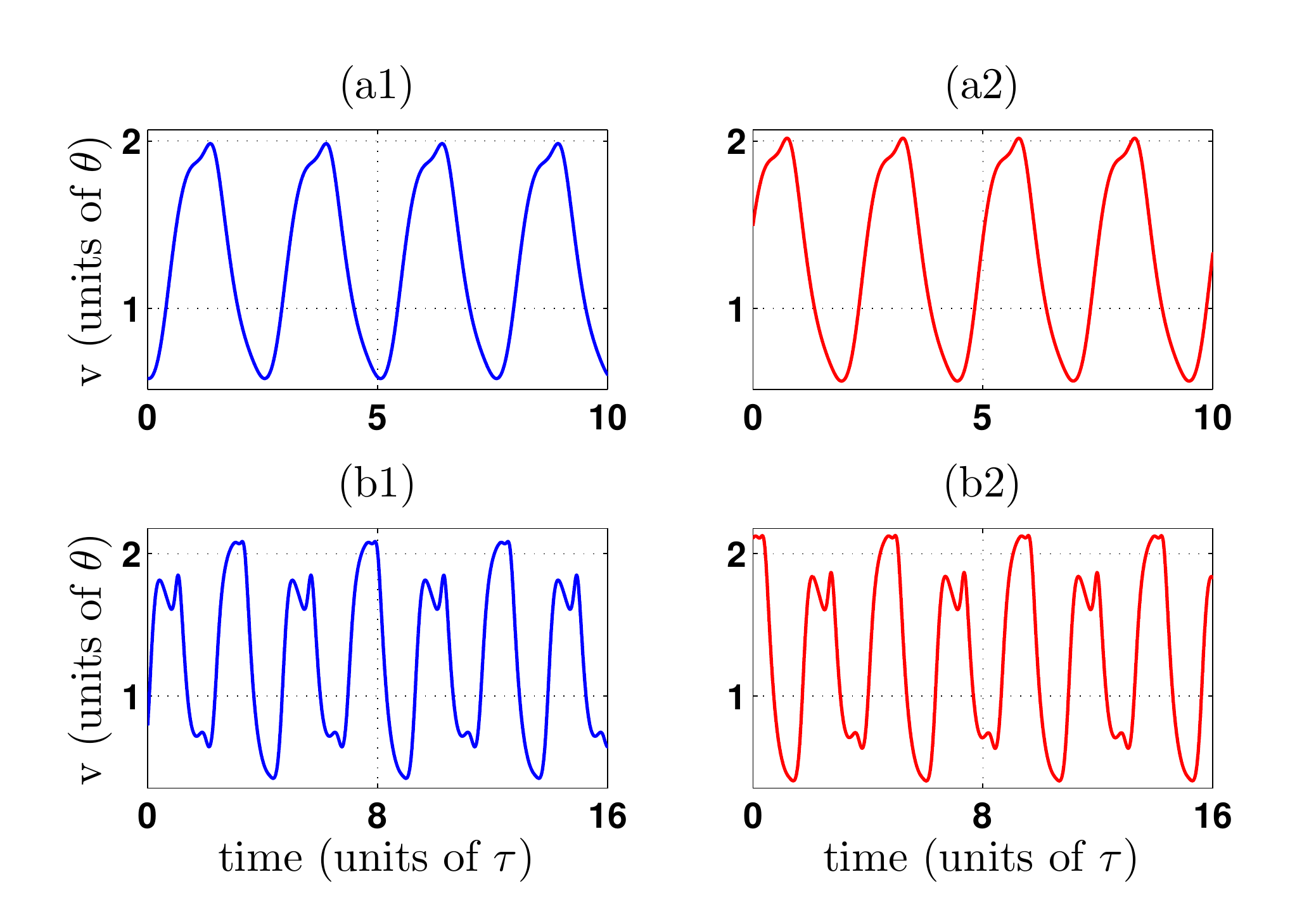} 
\caption{\label{fig:Series1}Exemplary temporal evolutions of MG model,  simulations (left and blue), and
  experiments (right and red). Parameter values: $\Gamma=3$ (a) and  $\Gamma=5$ (b).}
\end{figure}

\begin{figure}
\includegraphics[width=0.98\columnwidth]{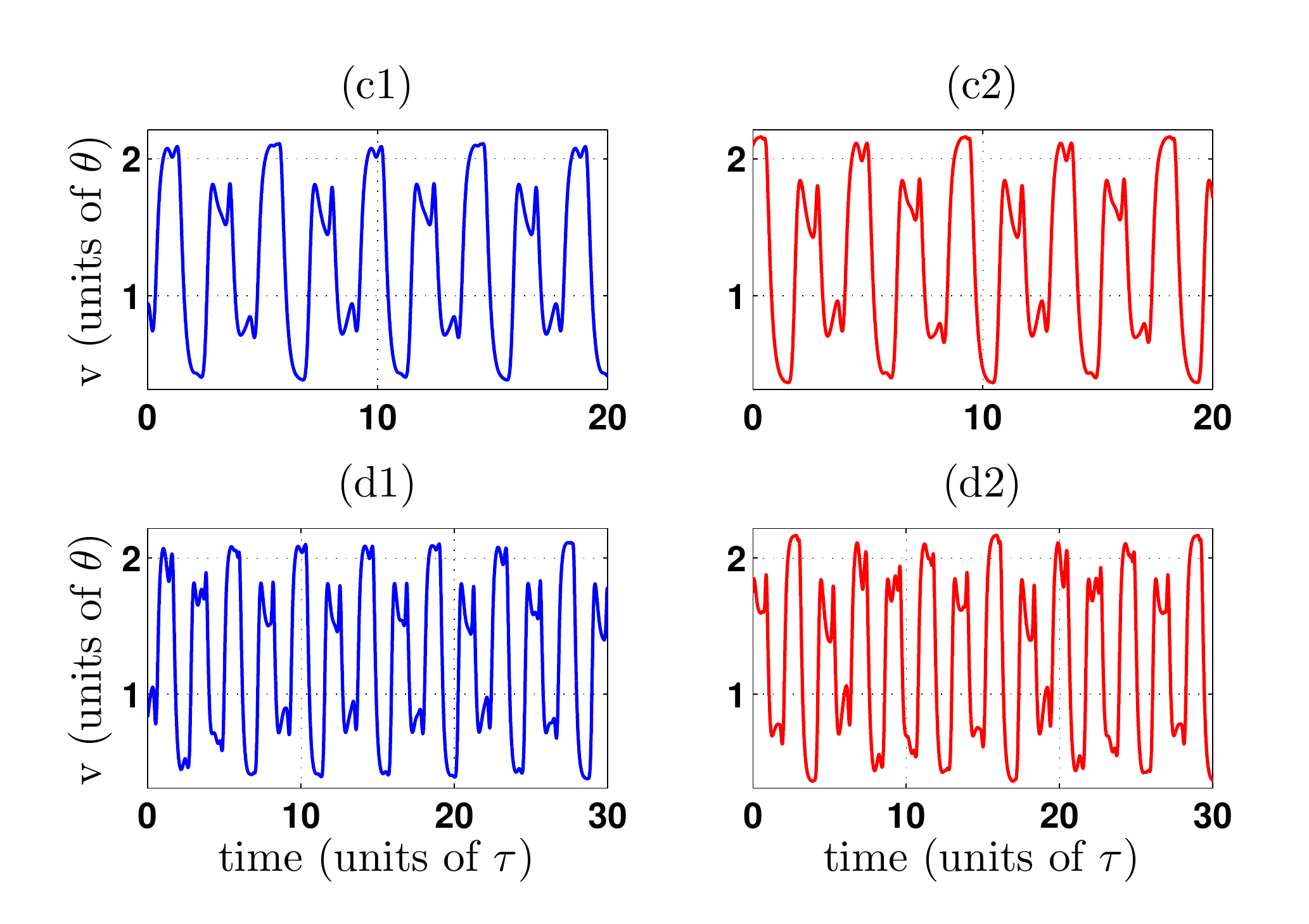}\caption{\label{fig:Series2}
Exemplary temporal evolutions of MG model,  simulations (left and blue), and experimental
data (right and red). Parameter values: 	$\Gamma=7$ (c) and $\Gamma=8.5$ (d).}
\end{figure}

Some examples of  time series with different $\Gamma$ values are shown in Fig. \ref{fig:Series1} and \ref{fig:Series2}, 
corresponding to the values indicated by  vertical dashed lines in Figs.~\ref{fig:D-TBif} and \ref{fig:E-TBif}. 
These time series also show a great deal of concordance between experimental and simulated data.
The first presented waveforms  ($a1$ and $a2$) correspond to a simple
oscillation with only one peak per period. Increasing the time delay, $\Gamma$,
after new peak appearances and a bifurcation, a more complex waveform, with longer period is observed ($b1$ and $b2$).
With additional increases in the time delay longer periods appear as shown in $c1$ and $c2$. Finally, 
chaotic behavior, in agreement with the previous works \cite{mackey1977oscillation,namajunas1995electronic,junges2012intricate}, 
is seen in $d1$ and $d2$.

\section{Conclusion and perspectives}
\label{sec:con}

The Mackey-Glass model was simulated using an electrical analog. A novel approach 
was proposed to implement the  function and delay blocks exhibiting several advantages
compared with previous implementations. An exact  equation for the
electric evolution can be easily written thanks to the exact transfer function  of the delay block.
The parameter $n$ is exactly determined by means of multipliers and divisors.
The ordering of the blocks, which is usually ignored, played an important 
role in getting the most out of the delay block, and allowed to suppress a most of the electrical noise in the experimental data.

The results of numerical simulations of the the original equation were compared with the effective (discrete) equation
and with the experimental data. According to the value of the time delay  the system exhibits a wide variety of behaviors including
fixed, periodic waveform with different numbers of peaks per period and, finally, aperiodic or chaotic solutions.
The great agreement between experimental and numerical results suggests that can deeper understanding of the dynamics of MG model can be obtained using
our implementation. In addition,  this approach could be extended to other time-delayed systems.

% 
% 
% 
% \appendices
% \section{Proof of the First Zonklar Equation}
% Appendix one text goes here.
% 
% % you can choose not to have a title for an appendix
% % if you want by leaving the argument blank
% \section{}
% Appendix two text goes here.
% 

% use section* for acknowledgement
\section*{Acknowledgment}

The authors would like to thank CSIC (UdelaR) and PEDECIBA (Uruguay).

% Can use something like this to put references on a page
% by themselves when using endfloat and the captionsoff option.
\ifCLASSOPTIONcaptionsoff
  \newpage
\fi

% trigger a \newpage just before the given reference
% number - used to balance the columns on the last page
% adjust value as needed - may need to be readjusted if
% the document is modified later
%\IEEEtriggeratref{8}
% The "triggered" command can be changed if desired:
%\IEEEtriggercmd{\enlargethispage{-5in}}

% references section

% can use a bibliography generated by BibTeX as a .bbl file
% BibTeX documentation can be easily obtained at:
% http://www.ctan.org/tex-archive/biblio/bibtex/contrib/doc/
% The IEEEtran BibTeX style support page is at:
% http://www.michaelshell.org/tex/ieeetran/bibtex/
\bibliographystyle{IEEEtran}

\bibliography{/home/marti/Dropbox/bibtex/mybib}

\begin{IEEEbiography}[{\includegraphics[width=1in,height=1.25in,clip,keepaspectratio]{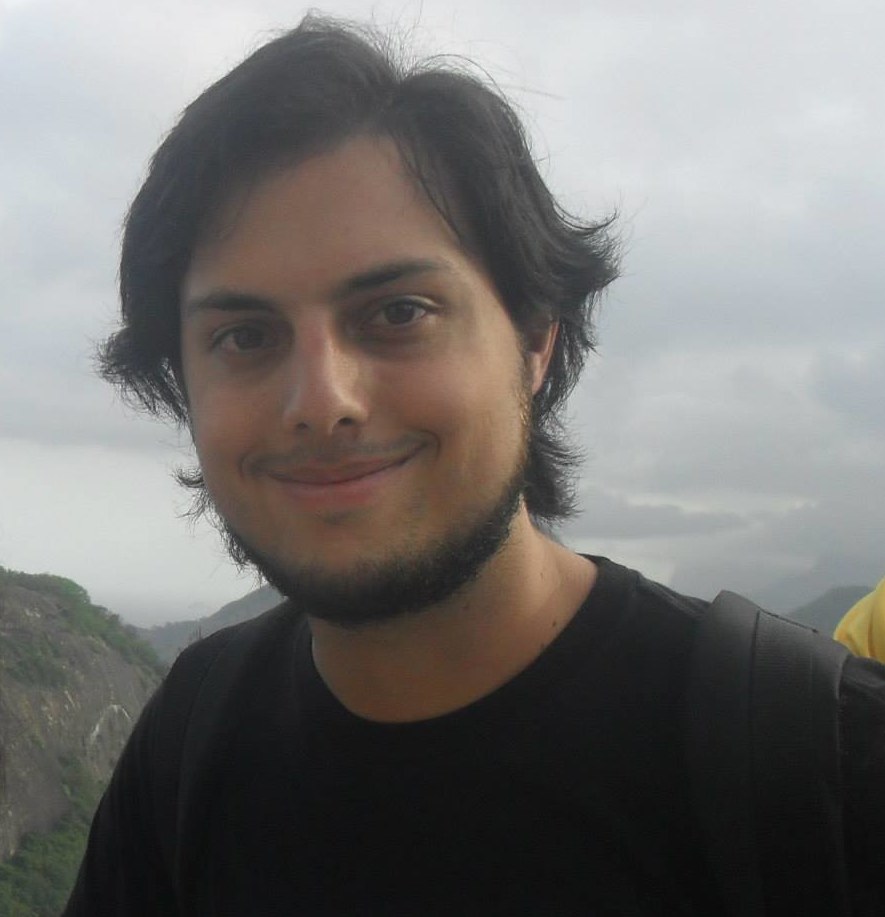}}]{Pablo Amil}
Pablo Amil received  his bachelor degree from the University of the Republic, Uruguay, in 2012. Where he is currently finishing a master degree.
His current research include nonlinear dynamics, and musical audio synthesis.
\end{IEEEbiography}

\begin{IEEEbiography}
[{\includegraphics[width=1in,height=1.25in,clip,keepaspectratio]{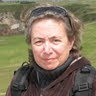}}]{Cecilia Cabeza}
Cecilia Cabeza is associate professor in Physics at the Republic University (Uruguay). Her research interests have 
focused on nonlineal physics, chaos, fluids, and teaching of experimental physics.  Cecilia holds a doctorate cotutelle from the Republic University (Uruguay) 
and the Paris VII University (France)  in 2000.

\end{IEEEbiography}

% insert where needed to balance the two columns on the last page with
% biographies
%\newpage

\begin{IEEEbiography}
[{\includegraphics[width=1in,height=1.25in,clip,keepaspectratio]{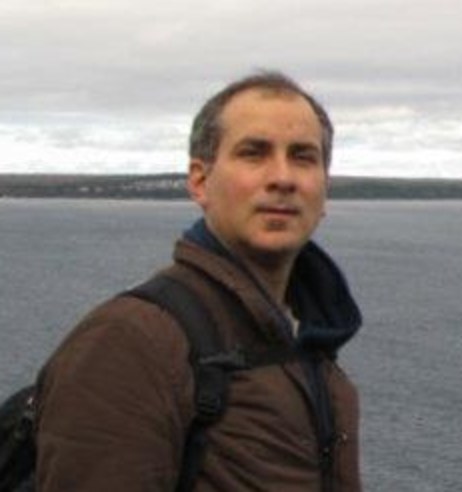}}]{Arturo C. Mart\'{\i}}
Arturo Marti is associate professor in Physics at the Republic University (Uruguay).
His research interests have centred on traditional academic topics as chaos, turbulence,  and complex networks but also include
science popularization and teaching of  physics using everyday tools.
Arturo obtained his degree in 1992 from the Republic University (Uruguay) and his PhD from Barcelona University  (Spain) in 1997.
\end{IEEEbiography}

% You can push biographies down or up by placing
% a \vfill before or after them. The appropriate
% use of \vfill depends on what kind of text is
% on the last page and whether or not the columns
% are being equalized.

%\vfill

% Can be used to pull up biographies so that the bottom of the last one
% is flush with the other column.
%\enlargethispage{-5in}

% that's all folks
\end{document}